# Investigation of variation in fluorescence intensity from rhodamine 6G Dye

S. M. Iftiquar[a,b*] and H. Zilay[c]

[a] SPMS, Nanyang Technological University, 21 Nanyang Link, Singapore 637371

[b] College of Information and Communication Engineering, Sungkyunkwan University, Suwon, S Korea

[c] Ajou University, 206 World cup-ro, Woncheon Dong, Yeongtong Gu, Suwon, South Korea

Abstract

Variation in fluorescence intensity from rhodamine 6G dye was investigated. A small volume of dye solution was optically excited with a 400 µW, 532nm wavelength cw-laser light. The dye was dissolved in methanol and glycerol for a concentration of 10mg/ml. With the optical excitation, initially the fluorescence intensity was observed to rise, and then it decayed, along with a steady shift of fluorescence peak from 562 nm to 543 nm. The observation of initial enhancement in fluorescence from start to 7 minutes of excitation, can partly be due to the low excitation power, therefore slower rate of change of fluorescence intensity with time. Simulation studies indicate that the photo bleaching was taking place from all the energy states of the dye molecules, which is an extension of the concept that the photo bleaching takes place at the excited triplet state whereas the fluorescence takes place due to transition between ground and excited singlet states. A steady shift in fluorescence peak position, from 562 nm to 543 nm, was observed during the fluorescence life of the dye, at a rate of 0.0113 nm/s during fluorescence enhancement and 0.026 nm/s during photo bleaching.

* Corresponding Author: smiftiquar@gmail.com

## 1. INTRODUCTION

Variation in fluorescence intensity with time is commonly observed in almost all dye materials. Rhodamine 6G (Rh6G) is one of them and also one of the most popular dye materials, and is relatively stable [1]. It has been used as a therapeutic mediator [2], fluorescent marker [3], in labeling biological molecules [4], it was also used in a tunable dye laser [5] as a laser dye, in dye sensitized solar cell [6], heat sensor [7], probing molecular environment [8] etc. In all these one noticeable characteristic feature is that the wavelength of emitted fluorescence radiation is almost always longer than the excitation wavelength [9].

In addition to that it was also noticed that the fluorescence intensity decays with a prolonged optical excitation, known as photo bleaching. In the photo bleaching the number of fluorescent molecule reduces with time and it happens due to a damage to the molecules ability to fluoresce. The characteristics of the photo bleaching or fluorescence decay, depends on various parameters like the type of dye molecule, intensity of the optical excitation, presence of oxygen [10], presence or absence of lasing, chemical environment [11] or solvent [12] etc.

According to Jablonski energy diagram, the whole process of fluorescence and photo bleaching can be described with the help of singlet triplet energy states [13]. The molecules in the ground state singlet state is excited to an excited singled state. From the excited state it can decay by fluorescence to the ground state or it can be transferred to excited triplet energy state. The molecules in the triplet state do not give fluorescence and it was considered as the state from which the photo bleaching happens fast. Generally the photo bleaching or the decay characteristics of fluorescence intensity is closely similar to an exponential decay [13]. Although the fluorescence decay was investigated for a long time, yet the investigations were either with a flow of dye solution or investigating a large volume of dye solution where the all the dye molecules were not illuminated continuously during the period of investigation.

This may lead to some missing information about the fluorescence characteristics when all the investigating molecules are isolated and yet uniformly illuminated during the period of investigation. For example, it may be possible that the characteristic variation in fluorescence intensity with time can be different from that reported in the literature. One new result that we obtained in our investigation is that the fluorescence intensity increases at the initial stage of optical excitation then it decays. This initial enhancement in fluorescence intensity has not been reported so far.

Therefore we investigated a small amount of dye solution in isolation, in the form of an isolated and levitated micro drop, and through a continuous period of optical excitation. By optically exciting the Rh6G dye within the droplet for a prolonged period of time and observing the variation in the fluorescence intensity with time we explored the characteristic variation in fluorescence intensity with time. The experimental method used here is described below.

## 2. EXPERIMENTAL

The investigation was performed in a trapped and levitated liquid micro drop. Electrically charged microdrop was created with the help of an electro spray and the droplet was levitated in a modified Paul trap (MPT) in normal atmospheric pressure.

The micro drop is composed of Rh6G dye, dissolved in a methanol glycerol mixture. It emits yellow radiation when illuminated by 532 nm green laser light. Figure 1a shows experimental setup that consists of an electro spray system and the MPT [14, 15]. The system was under normal atmospheric pressure, but kept enclosed to reduce effect of ambient air convection. Rh6G glycerol solution was pumped into the electro spray needle at a ~ 8 μliter per minute, this was done only before formation of the droplet, and the flow was stopped when a micro drop was formed. The droplet was formed as follows. A DC field of ~ 5 kV was applied to the electro spray needle, it created a spray of charged liquid drops when the liquid flows. These droplets pass through the MPT, few of them gets confined within the MPT. After that we selectively trap single droplet in the MPT and investigated its fluorescence.

For investigating the fluorescence, the micro drop was optically pumped with a 532 nm wavelength continuous wave (cw) laser beam. A part of the incident light is absorbed by the Rh6G within the droplet, rest are scattered and transmitted. The diameter of the droplet was kept smaller that the beam cross section of exciting light. When the Rh6G molecules absorbs light, it get energetically excited and then emits a characteristic fluorescent light. The light coming from the droplet was recorded in an optical spectrometer and image of the droplet was separately recorded in a CCD camera, placing them perpendicular to the line of path of the pump beam. We used optical fiber coupled Ocean Optics spectrometer, whose resolution can change for larger core optical fiber; with such fiber effective detection sensitivity of becomes better. With a fiber of 50 μm core diameter the detector resolution is 2.9 nm while with 600 μm fiber the resolution becomes 31.0 nm, whereas in the second case detector's effective sensitivity becomes higher than that with the 50 μm fiber, because more light enters the detector.

The MPT consists of four copper electrodes, a ~ 30 Volt DC bias was applied to the external pair of electrodes while a ~ 2 kV$_{p-p}$ voltage with ~ 600 Hz frequency electric field was applied to the inner pair of electrodes, here V$_{p-p}$ is peak to peak voltage. The inner electrodes are separated by 2 mm air gap, through which the droplet is optically pumped and the emitted radiation from droplet is collected. Figure 1b shows top-view of the trap for pumping and detection of trapped micro drop, while Fig. 1c shows image of a trapped droplet.

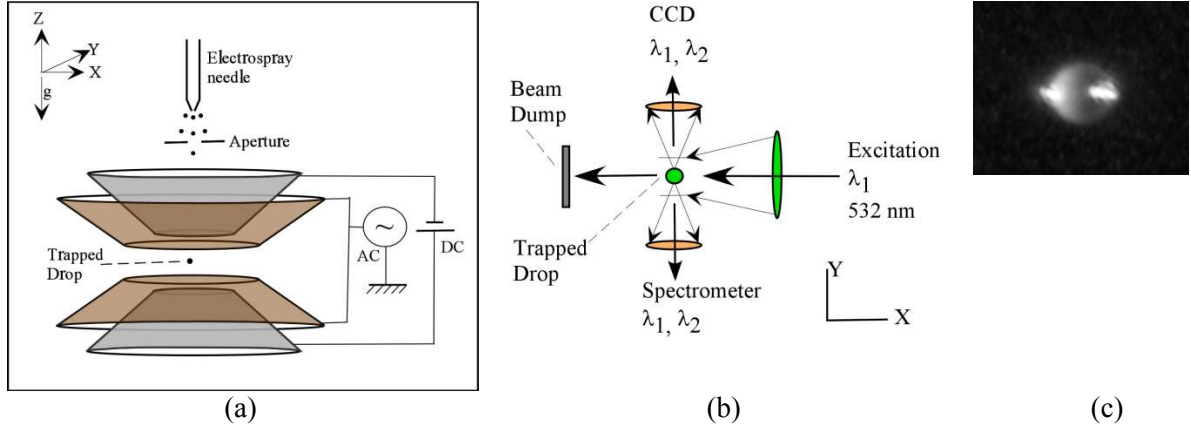

(a) (b) (c)

**Fig. 1.** (a) Schematic diagram of the electrodes for the modified Paul trap. (b) Schematic diagram of pumping and detection light. (c) A typical CCD image of a levitated micro drop in the modified Paul trap, of size 22.6 μm, observed with a 532 nm optical filter.

## 3. RESULTS

The detected spectra of the light coming from the micro drop contain part of the incident light reflected from the droplet and fluorescence from Rh6G. Figure 2 shows typical spectra, here the fluorescence intensity changes due to changing intensity of the excitation light.

Figure 2 shows a set of fluorescence spectra from Rh6G dye as well as the excitation light. With an increase in pumping power, the fluorescence spectra within 540 ~ 640 nm wavelength range, increases. However, there are small and sharp peaks visible at the uppermost most trace corresponding to whispering gallery modes (WGM) of resonance from the droplet [16]. The mutual separations among the WGM modes are expected to depend upon several factors [16]. However, in the present investigation we like to observe the characteristic intensity of fluorescence and photo bleaching of the Rh6G dye due to a continuous illumination. So 600 μm fiber was used for a better detection of weak signal, because of which the WGM will not be clearly visible.

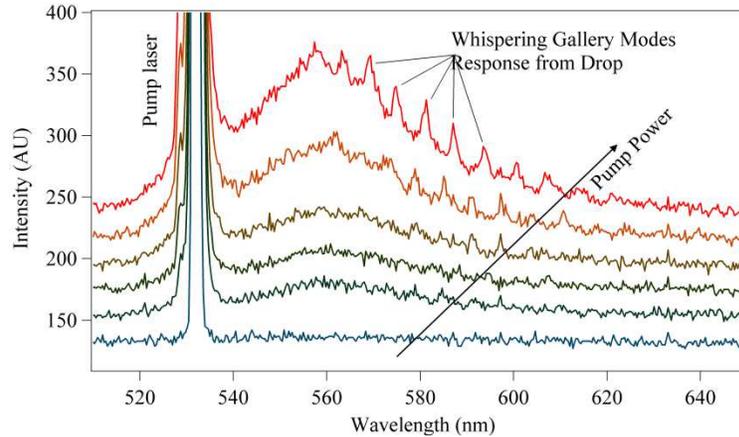

**Fig. 2.** Optical spectra of light collected by the spectrometer. The label 'pump laser' indicates the cw 532 nm pump laser that gets reflected from the surface of the droplet. Rest of the spectra for wavelength larger than 540 nm corresponds to fluorescence from the Rh6G dye. The lowest and highest spectra corresponds to 0.1 and 400 μWatt of pump light.

In order to record time dependence of fluorescence, the trapped drop was first stabilized within the MPT, then the optical excitation started. The fluorescence spectra were continuously recorded at an interval of 200ms, for about 1000 seconds. It was noticed that the fluorescence intensity from the Rh6G steadily increased until 408 seconds and then started decreasing. In order to see the trend clearly, the rising and falling spectra are displayed separately in Fig. 3(a) and (b) respectively. Here it can also be noticed that the peak of the fluorescence intensity is slowly shifting towards shorter wavelength.

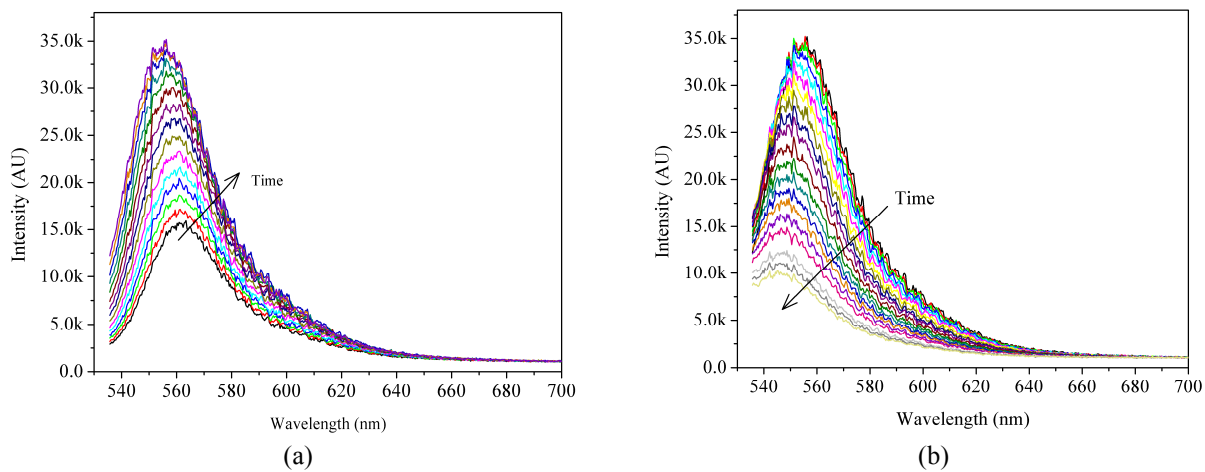

(a)　　　　　　　　　　　　　　　(b)

**Fig.3.** Fluorescence spectra for 13 micro meter diameter droplet. With Rh6G concentration of 10mg/ml and the spectra were recorded at 200 ms interval. (a) The enhancement in the fluorescence spectra with time at the initial stage when fluorescence intensity was increasing and reached a maximum. (b) The decay in intensity of the fluorescence spectra after reaching the maximum. Here the intensity started to fall.

Total fluorescence intensity (TFI) was obtained by integrating the area under the fluorescence spectra. Variation in the TFI with time is shown in Fig. 4(a). Figure 4(b) shows the change in peak position of the fluorescence spectra with time. The Fig. 4(a) clearly shows that the TFI increases for the first 7 minutes excitation, afterwards it decays. The time dependant intensity profiles looks like a distorted Gaussian type. A Gaussian fitting of the data, not shown here, gives a very good match with two deconvoluted Gaussian curves, having different peak height and width but with a same peak position. But such a Gaussian feature cannot be explained based on Jablonski energy diagram nor with the first order differential equation that described the transition of molecules from one energy level to another. Therefore the Gaussian feature of the spectra will not be analyzed. Rather we will use first order differential equations to describe all the possible transitions and solve then to obtain the dynamics of population density in the E2S state, from which the fluorescence occurs, and detected by the spectrometer. Here E2S is singlet second excited state.

The Fig. 4(b) shows that the peak positions of the fluorescence spectra changed from longer to a shorter wavelength (blue shifted), from 562 nm to 543 nm in 1000 seconds. The initial blue shift of the peak position or during the enhancement of the fluorescence, the rate of change of blue shift is smaller, 0.0113 nm/sec. On the other hand, after the maximum fluorescence, when the overall fluorescence starts decreasing, the rate of change of blue shift was observed to be 0.026 nm/sec.

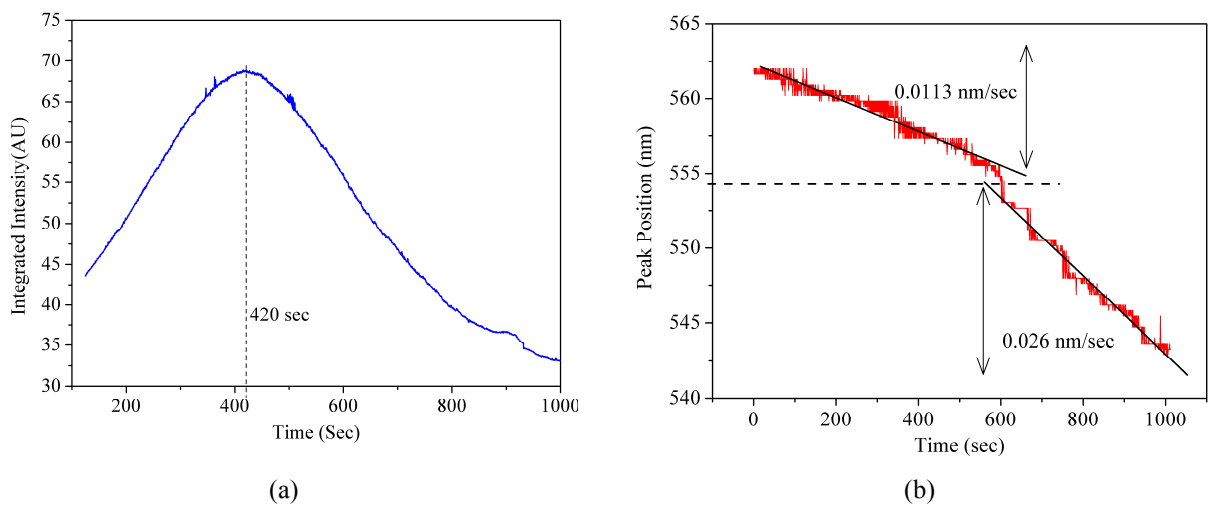

(a)           (b)

**Fig. 4.** (a) Variation in integrated fluorescence intensity with time. (b) Variation in peak position of the fluorescence spectra with time.

# 4. DISCUSSIONS

Rhodamine 6G is a popular fluorescent dye material, used in various applications. Most of the applications indicate that the fluorescence decays with time [5, 12, 17-22] due to photo bleaching.

## 4.1. PHOTO BLEACHING

The photo bleaching can be described with the help of Jablonski energy band diagram (JED). Here we use JED that is more suitable for our experiment, and is shown in Fig. 5.

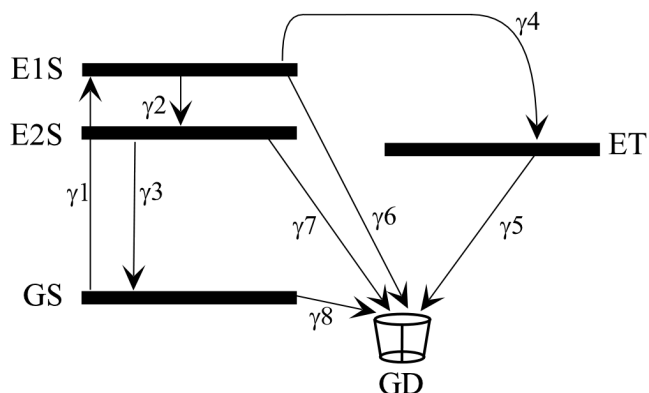

**Fig. 5.** Jablonski energy band diagram for the Rh6G molecules. GS – singlet ground state, E1S – singlet first excited state, E2S – singlet second excited state, ET – triplet excited state, GD – ground dump state or damaged molecular state, from where absorption-fluorescence does not occur. The $\gamma$'s are excitation rates and the number 1~8 correspond to the energy states as shown in the figure.

Before the optical excitations, the molecules are assumed to be initially at the singlet ground states (GS). With laser light the molecules get excited to E1S state, where it will thermalize to a relatively lower energy singlet excited state (E2S). From the E2S state it can de-excite to the GS or get transferred to the excited triplet state (ET), or move to the ground dump state thereby losing its ability to give fluorescence any more. The transition to the GD state is possible from all the energy state. From this ET state the molecules can decay to a permanently damaged or bleached state or it can come back to E2S state. The transition from the E2S state to the GS state leads to the observed fluorescence. It was suggested that lifetime of the triplet excited state is longer than the singlet excited state [5]. The molecules reaching the ET will not give fluorescence unless they are transferred back to the E1S or E2S state [13, 23]. Fluorescence quenching or spectral hole burning experiments with single molecules indicate that there can be a sudden dark state of the molecule, when it will not give fluorescence, probably because they get transferred to the ET state [13, 23]. The fluorescence can be observed again after some time, probably because the molecules in the ET state gets transferred back to the E1S or E2S state that result in the re-occurrence of fluorescence. In the JED, as the vertical energy difference between the ET and E2S is smaller than the GS, so we assume that most of the desired transitions from ET will be thermally activated to the E2S state.

The above mechanism is what we propose as best suited for our observation. The observation of slower or faster photo bleaching is expected to depend upon various parameters, few of them are photon flux or excitation power [3, 21, 24, 25] (in Rh6G, fluorescein dextran, coumarin dextran, and Indo-1 pentopotassium salt), temperature [3], presence of oxygen [13] etc. At a higher photon flux a faster decay is expected, similarly at a higher temperature, the decay may also be faster [3]. We used a low power excitation laser. This helps to observe the fluorescence dynamics more easily. The population density in the E1S, E2S and ET states are expected to increase with increased intensity of excitation light [25], leading to a faster process, that may not be convenient to explore the details of time evolution of fluorescence intensity.

It is also suggested that a large volume of Rh6G dye will degrade faster during illumination with a smaller light of beam, so a smaller volume dye may have a slower rate of photo bleaching because of better uniformity in illumination [5]. For that purpose the micro drop is a good way to use a small volume of dye.

## 4.2. THEORETICAL MODEL

It is generally accepted that fluorescence intensity ($I_{FL}$) decays exponentially [1], and can be expressed as a single exponentially decaying function [13, 26, 27]:

$$I_{FL} = A + Be^{-kt}$$

Here A represents background fluorescence or the fluorescence that does not decay with time, B is amplitude of fluorescence decay, k is decay constant, t is time. This single exponential decay may be primarily induced by oxygen [13]. However, in general a deviation from single exponential decay is expected [13, 28, 29].

If the fluorescence decay follows the single exponential relation, then a semilog plot should give a straight line fit [28, 29]. Figure 6 shows the log of intensity plotted with time. In this figure, a linear best fit trace is equivalent to a single exponential trace with time. The curves in Fig. 6(a), (b) indicates that the fluorescence enhancement and decay does not follow single exponential relation as the enhancement and the decay profiles are deviated from the straight line fit.

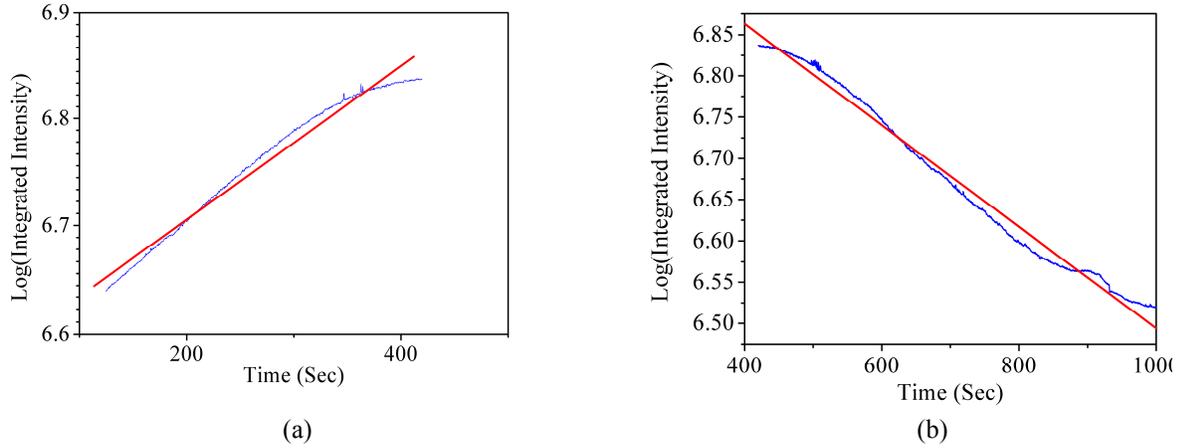

**Fig. 6.** Log of fluorescent intensity plotted with time. (a) During initial rise in fluorescence, (b) during fluorescence decay. The blue colored continuous trace is experimental data while the straight lines are linear best fits.

Mathematically the dynamics of the density of molecules at the various energy states can be expressed as follows:

$$\frac{dGS}{dt} = -\gamma_1 GS + \gamma_1 E1S + \gamma_3 E2S - \gamma_8 GS \qquad (1)$$

$$\frac{dE1S}{dt} = \gamma_1 GS + \gamma_4 ET - \gamma_1 E1S - \gamma_2 E1S - \gamma_4 E1S - \gamma_6 E1S \qquad (2)$$

$$\frac{dE2S}{dt} = \gamma_3 GS + \gamma_2 E1S - \gamma_3 E2S - \gamma_7 E2S \qquad (3)$$

$$\frac{dET}{dt} = \gamma_4 E1S - \gamma_4 ET - \gamma_5 ET - \gamma_4 E1S \qquad (4)$$

Here GS, E1S, E2S, ET are the number density of molecules in singlet ground, first excited, second excited and triplet excited states, whereas the γ's indicate transition constant and subscripts are used for different transitions, as indicated in the Jablonski energy diagram (Fig. 5 ).

We solved the first order differential equations numerically. As the detected fluorescence at ~600 nm wavelength corresponds to emission due to transition from E2S to GS, so the time variation of the E2S is shown in the Fig. 7. It shows that an initial rise in fluorescence intensity is expected due to a gradual build up of population in the E2S state. Then a decay in the fluorescence occurs due to a dominating photo bleaching. Here, the simulation results show that, the temporal profile has a longer tail than that observed in experiment. This may be because the effect of temperature that was not considered in the solution of the equations (1)~(4). On the other hand, in the experiment the rise in temperature is expected, making the decay faster in experimentally observed results than that in simulation. The decay rate, which does not exactly follow a single exponential relation, may be because the decay happens from multiple energy states, as a single exponential decay corresponds to purely single exponential decay, but when the decay happens from multiple energy levels with different decay constants then the decay is expected not to follow a single exponential curve.

The initial rise in fluorescence can be clearly observed if intensity of the excitation radiation is low. In our experiment the excitation intensity was 815 mW/cm$^2$ intensity, which is much smaller than the fluorescence degradations reported in the literature.

In this situation, for the initial 7 minutes the fluorescence intensity was observed to rise with time. The theoretical simulation shows that with a smaller excitation rate the Rh6G molecules at the GS will take time to be in the excited states. During this time the rise in the fluorescence is visible. Furthermore, the liquid micro drop can work as a laser cavity with whispering gallery mode of resonance. Because of which the stimulated emission from the E2S excited state is possible due to Doppler shift of the resonance for molecules (due to thermal motion) within the liquid. Therefore, the transition from E2S singlet to triplet state is expected to be reduced further. This can also lead to a reduction in photo bleaching rate.

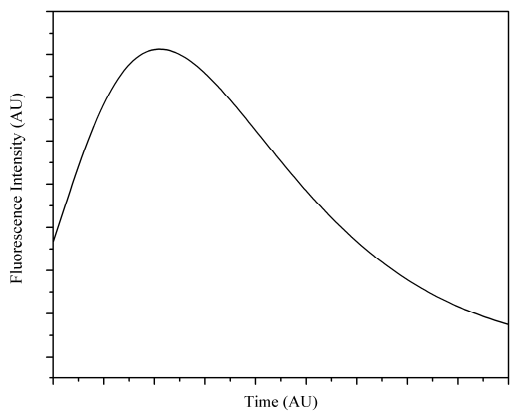

**Fig. 7.** Simulated time evolution of the E2S density of states.

## 4.3. CONCENTRATION

The fluorescence peak wavelength may depend upon concentration of fluorescent molecule. For example, at a lower concentration the peak position can be 550 nm and at a higher concentration it can be at 602 nm [30]. Based on this effective concentration of the Rh6G may qualitatively be estimated. Fig. 4(b) shows that the fluorescence peak continuously shifting from 562 nm to 543 nm, although at the first half of the experiment, the rate of change was lower at 0.0113 nm/s than that at the later part of fluorescence, when it was 0.026 nm/s. Figure 4(a) indicates that there is an enhancement in fluorescence intensity but at the same time the fluorescent peak continued shifting towards shorter wavelength. Therefore, it can be concluded that the rise in fluorescent intensity is primarily because more and more Rh6G molecules were excited to the excited singlet state, as well as the photo bleaching continued. The photo bleaching reduces effective concentration of Rh6G, because of which the fluorescent peak continued to move towards shorter wavelength. It is interesting to note that in a dilute form, the rhodamine based dye have a low quantum yield for transition to the triplet state [8, 31]. Fig. 4(a) shows that the decay in fluorescence happens at a slower rate for time > 10 min, and the curve becomes flatter with time. As the decay rate may possibly tend to slow down due to reduced effective concentration, so the curve becomes flatter with time.

## 4.4. TEMPERATURE

It was reported that at a higher temperature the photo-bleaching happens at a faster rate [3], probably because at a higher temperature the change in molecular structure is faster. Here, in this experiment, as the droplet was continuously illuminated with the laser, its temperature is expected to rise with time as more and more light was absorbed by the Rh6G. At a higher temperature, the photo-bleaching can happen faster, therefore the observed rate of change of the peak position of the fluorescence spectra also became higher with a passage of time. Although we observed only two distinct rate of change of the fluorescent peak position. As the droplet was isolated and levitated in air, so the absorbed heat is expected to stay confined to the droplet for longer. Therefore, even though the photo-bleaching leads to less number of Rh6G molecules to absorb incident light, yet the already existing temperature leads to its steady and faster photo-bleaching.

This faster photo-bleaching can also be reflected in the Fig. 4(a), after 7 minutes of fluorescence. Theoretically, or numerical simulation indicates that the rise in fluorescence will be much faster than the fall in fluorescence, leading to an ideal fluorescence intensity curve that rises sharply at the beginning and then falling in intensity at a much slower rate, as shown in Fig. 6. But in our experiment (Fig. 4(a)) the intensity curve looks almost symmetrical around its peak. Faster photo bleaching at a higher temperature and a slower one at a reduced concentration may mutually reduce each other's effect.

Various other mechanisms of photo bleaching have been suggested. Presence of oxygen can be one of them. Due to reaction of oxygen with the fluorescent material, its ability to fluoresce degrades [10]. However, other report suggests that such oxygen dependant photo bleaching is not very convincing [13].

## 5. CONCLUSIONS

The fluorescence experiment with Rh6G dye in a levitated microdrop and with a low pump power can lead to observation of initial increase in fluorescence and then its decay. The observation of initial rise in fluorescence is possible if rate of optical excitation is low. Due to the deviation of fluorescence decay from a single exponential relation, it can be concluded that the decay is happening from multiple energy levels. Furthermore, temperature is also expected to have a role in the photo-beaching; higher the temperature, faster is the photo bleaching. This result indicates that the Rh6G can give fluorescence for longer period if low excitation power is used along with a low temperature. The decay obtained in simulation is relatively slower than that was observed experimentally, probably due to temperature induced enhancement in photo-bleaching that was taking place in the experiment.